# Gaussian Mixture Model Based Distributionally Robust Optimal Power Flow With CVaR Constraints

Lei You, *Student Member, IEEE*, Hui Ma, *Senior Member, IEEE*, Tapan Kumar Saha, *Fellow, IEEE*, Gang Liu, *Senior Member, IEEE*

*Abstract*—This paper proposes a distributionally robust optimal power flow (OPF) model for transmission grids with wind power generation. The model uses the conditional value-at-risk (CVaR) constraints to control the reserve and branch flow limit violations caused by wind power forecast errors. Meanwhile, the Gaussian mixture model (GMM) is integrated into the CVaR constraints to guard against the non-Gaussian forecast error distributions. Unlike the previous studies considering the GMM with fixed parameters, this paper allows the GMM parameters to be variable within some credible regions and develops a data-driven GMM-based ambiguity set to achieve the distributional robustness. Also, rather than using the traditional sample-based approximation of CVaR with high computational burden, this paper designs a scalable cutting-plane algorithm to handle the distributionally robust CVaR constraints. Case studies on the IEEE 2736-bus system show the effectiveness and scalability of the proposed OPF model.

## Nomenclature

**Sets and Indices**

| | |
|---|---|
| $b, g, l, i, m, n, k$ | Index for load, traditional unit, transmission line, wind farm, Gaussian component, bootstrap resample and random constraint, respectively. |
| $\mathcal{B}, \mathcal{G}, \mathcal{L}, \mathcal{W}$ | Sets of buses, traditional units, transmission lines and wind farms, respectively |

**Parameters**

| | |
|---|---|
| $\hat{w}_i$ | Wind power central forecast at wind farm $i$. |
| $\boldsymbol{H}^W_{(l,i)}, \boldsymbol{H}^G_{(l,g)}, \boldsymbol{H}^D_{(l,b)}$ | Shift distribution factor of wind farm $i$, unit $g$ and load $b$ to line $l$, respectively. |
| $c_g^F(\cdot)$ | Quadratic fuel cost function of unit $g$. |
| $c_g^{up}/c_g^{dn}$ | Up-/down-ward reserve price of unit $g$. |
| $d_b$ | Power demand of load $b$. |
| $\bar{p}_g/\underline{p}_g$ | Maximum/minimum capacity of unit $g$. |
| $\bar{R}_g^{up}/\bar{R}_g^{dn}$ | Maximum up-/down-ward reserve of unit $g$. |
| $\bar{f}_l$ | Transmission limit of line $l$. |
| $M, N, K, J, W$ | Number of Gaussian components, bootstrap resamples, random constraints, historical data and wind farms, respectively. |

**Variables**

L. You, H. Ma and T. K. Saha are within the School of ITEE, University of Queensland, Brisbane, Queensland 4072, Australia.
G. Liu is within the School of Electric Power, South China University of Technology, Guangzhou, Guangdong 510641, China.

| | |
|---|---|
| $\tilde{\xi}_i$ | Random wind power forecast error at farm $i$. |
| $\alpha_g$ | Participation factor of unit $g$. |
| $\hat{p}_g$ | Nominal output of unit $g$. |
| $R_g^{up}/R_g^{dn}$ | Up-/down-ward reserve provided by unit $g$. |
| $\hat{f}_l$ | Nominal power flow on line $l$. |
| $\pi_m, \boldsymbol{\mu}_m, \boldsymbol{\Sigma}_m$ | Mixing weight, mean and covariance of $m_{th}$ gaussian component, respectively. |

## I. Introduction

The forecast errors of wind power can pose considerable risk to power system security. To handle the wind power forecast uncertainties, different decision-making approaches have been proposed, such as the stochastic optimization (SO) using discrete wind power scenarios [1] and the robust optimization (RO) leveraging the uncertainty sets of wind power [2].

The distributionally robust optimization (DRO) has been emerging as an intermediate between SO and RO. DRO assumes the true distribution of the forecast error lies in an ambiguity set and makes decisions to hedge against the worst-case distribution. The moment-based ambiguity set has been used for power system dispatch [3], [4]. However, such ambiguity set utilizes only the moment information of the forecast error and may be very conservative. Another type of ambiguity set leverages the statistical distance between distributions, such as the Kullback-Leibler (KL) distance and Wasserstein distance. In [5], a KL-based ambiguity set is used for the unit commitment. In [6], a Wasserstein-based ambiguity set is applied to the optimal power flow (OPF) problem. However, KL-DRO cannot be applied if the uncertainties have heavy tails [7], while the model size of Wasserstein-DRO grows heavily with the data size employed. In [8], [9], a new type of ambiguity set is also constructed based on the cumulative probability distributions of random variables, but the correlations between random variables cannot be captured.

The decision-making under uncertainty also often needs some risk management tools. Value-at-risk (VaR) and conditional value-at-risk (CVaR) are two popular tools and have been successfully applied to power system problems. Based on VaR, the chance constrained OPF and UC models have been widely used to ensure the power system security with specified probability levels [10], [11]. The use of VaR is intuitive since it limits the violation probability of a constraint. However, VaR neglects the aftermath if the violation occurs. Also, the VaR-based chance constraints are often non-linear and non-convex, and thus are computationally intractable.

Unlike VaR, CVaR is a coherent risk measure [12]. It is defined on the mean of the distributional tail exceeding VaR and can thus account for both the probability and magnitude of constraint violation. Besides, the direct optimization of CVaR is a convex problem and is thus more tractable. In power systems, CVaR has been used for demand-side management [13] and unit commitment optimization [14].

Overall, both VaR and CVaR have their rationality and the decision maker may prefer one over the other in different situations. This paper focuses on CVaR. More specific, this paper aims to address two issues in the application of CVaR.

First, the calculation of CVaR may be still not easy since it involves high-dimensional integral. Currently, the sample-based approximation of CVaR is often used [15], but it may cause a huge computational burden under large sample sizes. An analytical reformulation of CVaR is also provided in [15], but it is valid only when the uncertainties follow Gaussian distributions. In reality, the forecast error of wind power often follows non-Gaussian distributions [16]. Recently, Gaussian mixture model (GMM) has been used to fit non-Gaussian forecast errors [17], [18]. However, the CVaR under GMM still lacks an analytical form. It is also noteworthy that though GMM has been used for VaR-based power system scheduling [19], [20], there is no existing work applying GMM to CVaR-constrained power system dispatch problems.

Second, in practice, the precise probability distribution of the forecast error is often unavailable. It could be biased to use the distribution estimated directly from limited historical data or based on expert knowledge. To address this issue, the moment-based ambiguity set with the mean and covariance information has been used to consider the worst-case CVaR [21]. Since the utilization of only moment information may result in very conservative solutions, the unimodality knowledge is further included in [22] to reduce the conservatism. Besides, the Wasserstein-based ambiguity set has also been used for CVaR-constrained OPF [23], but the resulting model is not scalable since the model size is proportional to the number of historical data. Also for the CVaR-constrained OPF, [24] provides an alternative method by combining the sample-based CVaR approximation with an ambiguous discrete probability distribution. However, similar to [23], this method is still not scalable since its size increases heavily with the number of forecast error samples.

This paper proposes a new distributionally robust OPF model with CVaR constraints. A data-driven GMM-based ambiguity set is used to hedge against the ambiguous true distribution of wind power forecast error. A scalable solution approach without any sampling procedure is also developed for the proposed OPF model. The contributions of this paper include:

1) This paper extends the application of GMM in power system operational planning. First, previous research often uses GMM in VaR-constrained dispatch problems, while it is the first time for GMM to be applied to CVaR-constrained OPF. Second, the traditional GMM has fixed parameters (e.g., see [19], [20]), while the GMM in this paper is allowed to have uncertain parameters by adopting a data-driven GMM-based ambiguity set. The use of the proposed GMM is more robust to the distributional ambiguity of wind power forecast error.

2) A scalable cutting-plane-based methodology is used to solve the proposed OPF model. Unlike the traditional methods to handle CVaR, the proposed methodology does not rely on any samplings when evaluating the worst-case CVaR and enforcing the distributionally robust (DR)-CVaR constraints and thus avoids the associated high computational burden (see, e.g., [24]). By exploiting the special properties and structure of the proposed GMM-based ambiguity set, a method is designed to quickly evaluate the worst-case CVaR. Then, based on the convexity and sensitivities of CVaR, cutting-planes are used to enforce the DR-CVaR constraints. Finally, an iterative cutting-plane-based solution algorithm is developed for the proposed OPF model. The whole methodology demonstrates high computational efficiency even for very large-scale networks.

## II. CVaR Constrained Optimal Power Flow

### A. Optimal Power Flow (OPF) Formulation

As the response to wind power forecast errors, we assume the outputs of traditional units are adjusted based on the policy:

$$\Delta \tilde{p}_g = -\alpha_g \sum_{i \in \mathcal{W}} \tilde{\xi}_i, \forall g \in \mathcal{G} \quad (1)$$

where the total power imbalance is allocated among the traditional units according to the participation factors $\alpha_g$. If $\sum_g \alpha_g = 1$, the system-wide power balance is guaranteed.

With the above policy, the OPF dispatch with uncertain wind power can be modelled as (P1):

$$(\mathbf{P1}): \min_{\hat{\boldsymbol{p}}, \boldsymbol{R}^{up}, \boldsymbol{R}^{dn}, \boldsymbol{\alpha}, \hat{\boldsymbol{f}}} \sum_{g \in \mathcal{G}} \{c_g^F(\hat{p}_g) + c_g^{up} R_g^{up} + c_g^{dn} R_g^{dn}\} \quad (2a)$$

s.t. $0 \leq R_g^{up} \leq \overline{R}_g^{up}, \quad \forall g \in \mathcal{G} \quad (2b)$

$0 \leq R_g^{dn} \leq \overline{R}_g^{dn}, \quad \forall g \in \mathcal{G} \quad (2c)$

$\underline{p}_g + R_{gt}^{dn} \leq \hat{p}_g \leq \overline{p}_g - R_{gt}^{up}, \quad \forall g \in \mathcal{G} \quad (2d)$

$\sum_{g \in \mathcal{G}} \hat{p}_g + \sum_{i \in \mathcal{W}} \hat{w}_i = \sum_{b \in \mathcal{B}} d_b \quad (2e)$

$\sum_{g \in \mathcal{G}} \alpha_g = 1; \quad \alpha_g \geq 0, \quad \forall g \in \mathcal{G} \quad (2f)$

$-\alpha_g \sum_{i \in \mathcal{W}} \tilde{\xi}_i \leq R_g^{up}, \quad \forall g \in \mathcal{G} \quad (2g)$

$\alpha_g \sum_{i \in \mathcal{W}} \tilde{\xi}_i \leq R_g^{dn}, \quad \forall g \in \mathcal{G} \quad (2h)$

$\hat{f}_l = \sum_{g \in \mathcal{G}} \boldsymbol{H}^G_{(l,g)} \hat{p}_g + \sum_{i \in \mathcal{W}} \boldsymbol{H}^W_{(l,i)} \hat{w}_i - \sum_{b \in \mathcal{B}} \boldsymbol{H}^D_{(l,b)} d_b, \forall l \in \mathcal{L} \quad (2i)$

$\hat{f}_l + \sum_{i \in \mathcal{W}} (\boldsymbol{H}^W_{(l,i)} - \sum_{g \in \mathcal{G}} \boldsymbol{H}^G_{(l,g)} \alpha_g) \tilde{\xi}_i \leq \overline{f}_l, \quad \forall l \in \mathcal{L} \quad (2j)$

$\hat{f}_l + \sum_{i \in \mathcal{W}} (\boldsymbol{H}^W_{(l,i)} - \sum_{g \in \mathcal{G}} \boldsymbol{H}^G_{(l,g)} \alpha_g) \tilde{\xi}_i \geq -\overline{f}_l, \quad \forall l \in \mathcal{L} \quad (2k)$

where (2a) minimizes the sum of the fuel cost and reserve-procuring cost; (2b)-(2d) limit the available up- and down-ward reserves; (2e)-(2f) enforce system-wide power balance; (2g)-(2h) bound the deployment of up- and down-ward reserves; (2i) indicates the nominal branch flow while (2j)-(2k) represent the random branch flow under the wind power fluctuations.

In the following analysis, the random constraints (2g)-(2h) and (2j)-(2k) are all represented by the general form:

$$\boldsymbol{y}_k(\boldsymbol{x})^T \tilde{\boldsymbol{\xi}} \leq h_k(\boldsymbol{x}), \forall k = 1, \cdots, K \quad (3)$$

where $\boldsymbol{x}$ is the decision vector in (P1), $\boldsymbol{y}_k(\boldsymbol{x})$ and $h_k(\boldsymbol{x})$ are two affine transformations of $\boldsymbol{x}$, $\tilde{\boldsymbol{\xi}} = \left(\tilde{\xi}_i, \forall i \in \mathcal{W}\right)^T$, $K$ is the total number of the random constraints. In the following analysis, $\boldsymbol{y}_k(\boldsymbol{x})$ is re-expressed as $\boldsymbol{y}_k$ to simplify notation.

*B. Risk Measure: Conditional Value at Risk (CVaR)*

To control the violation risk of (3), we resort to CVaR. Given a risk level $\beta \in (0, 1)$, the CVaR of $\boldsymbol{y}_k^T \tilde{\boldsymbol{\xi}}$ is defined as [12]

$$\text{CVaR}_{1-\beta}\left(\boldsymbol{y}_k^T \tilde{\boldsymbol{\xi}}\right) = \min_{t \in \mathbb{R}} \left\{ t + \frac{1}{\beta} \mathbb{E}_{\mathbb{P}}\left[\boldsymbol{y}_k^T \tilde{\boldsymbol{\xi}} - t\right]^+ \right\} \quad (4)$$

where $\mathbb{E}_{\mathbb{P}}[\cdot]^+ = \mathbb{E}_{\mathbb{P}}[\max(\cdot, 0)]$.

In (4), CVaR is the conditional expectation of $\boldsymbol{y}_k^T \tilde{\boldsymbol{\xi}}$ exceeding the optimal solution of $t$ (i.e., the value-at-risk of $\boldsymbol{y}_k^T \tilde{\boldsymbol{\xi}}$ at the probability level $1 - \beta$). Then, we use the CVaR constraints in (5) to control the violation risk of (3).

$$\text{CVaR}_{1-\beta}\left(\boldsymbol{y}_k^T \tilde{\boldsymbol{\xi}}\right) \leq h_k(\boldsymbol{x}), \forall k = 1, \cdots, K \quad (5)$$

However, the use of (5) has the following two issues:

i) Usually, we only have limited historical data of $\tilde{\boldsymbol{\xi}}$ and it is hard to estimate the precise distribution of $\tilde{\boldsymbol{\xi}}$. It can be biased to estimate CVaR by assuming any parametric distribution of $\tilde{\boldsymbol{\xi}}$.

ii) CVaR lacks a general computational efficient form. Though (5) can be handled by a sample-based approximation, the approximation quality could be sensitive to the number of samples and the sampling technique, and the computational burden could be huge when the number of samples is large.

## III. Distributionally Robust Formulation

To address the aforementioned two issues in Section II-B, we propose a distributionally robust approach combining the CVaR constraints (5) with a GMM-based ambiguity set. This is based on the following considerations:

i) GMM is adopted due to its flexibility in modelling various types of multivariate non-Gaussian probability distributions.

ii) The GMM fitted on limited training data may deviate from the unknown true distribution of $\tilde{\boldsymbol{\xi}}$, so a GMM-based ambiguity set is used to account for a GMM with uncertain parameters and hedge against the distributional ambiguity of $\tilde{\boldsymbol{\xi}}$.

iii) Though the GMM-based CVaR still lacks an analytical form, the use of GMM makes it possible to apply a cutting-plane method to enforce the CVaR constraints (5) (see Section IV). Thus, the sample-based approximation of CVaR is avoided.

*A. Gaussian Mixture Model (GMM)*

The probability density function (PDF) of the wind power forecast error $\tilde{\boldsymbol{\xi}}$ is defined as the following GMM:

$$\text{PDF}\left(\tilde{\boldsymbol{\xi}}|\boldsymbol{\Psi}\right) = \sum_{m=1}^{M} \pi_m \phi(\boldsymbol{\mu}_m, \boldsymbol{\Sigma}_m) \quad (6)$$

where $\phi(\boldsymbol{\mu}_m, \boldsymbol{\Sigma}_m)$ is the multivariate Gaussian distribution with the mean $\boldsymbol{\mu}_m$ and covariance matrix $\boldsymbol{\Sigma}_m$, and it is the $m_{th}$ Gaussian component of the GMM; $(\pi_1, ..., \pi_M)$ are the mixing weights of the $M$ Gaussian components; $\boldsymbol{\Psi}$ is the vector form of $\{\pi_1, ..., \pi_M, \boldsymbol{\mu}_1, ..., \boldsymbol{\mu}_M, \boldsymbol{\Sigma}_1, ..., \boldsymbol{\Sigma}_M\}$.

Given the historical data of $\tilde{\boldsymbol{\xi}}$, the parameters $M$ and $\boldsymbol{\Psi}$ can be obtained by the maximum likelihood estimation (MLE) [25].

*B. Constructing Credible Regions for GMM Parameters*

Different to the traditional GMM with fixed parameters, we assign some GMM parameters with certain uncertainty and allow them to be variable within some credible regions. This yields an ambiguous GMM. The credible regions are as follows.

i) Weights $\boldsymbol{\pi}$: $(\pi_1, ..., \pi_M)$ for all GMM components:

$$\mathcal{C}^\pi = \left\{ \boldsymbol{\pi} \middle| \begin{array}{c} \sum_{m=1}^{M} \pi_m = 1, \pi_m \geq 0, \underline{\pi}_m^\delta \leq \pi_m \leq \bar{\pi}_m^\delta, \\ \forall m = 1, ..., M \end{array} \right\} \quad (7)$$

where $\delta$ is a confidence level (e.g., 0.95) and $[\bar{\pi}_m^\delta, \underline{\pi}_m^\delta]$ is the $\delta$-confidence interval of $\pi_m$.

ii) Mean $\boldsymbol{\mu}_m$ for each GMM component $m = 1, ..., M$:

$$\mathcal{C}_m^\mu = \{\boldsymbol{\mu}_m | \ (\boldsymbol{\mu}_m - \hat{\boldsymbol{\mu}}_m)^T \boldsymbol{\Lambda}_m^{-1} (\boldsymbol{\mu}_m - \hat{\boldsymbol{\mu}}_m) \leq \gamma_m^\mu \ \} \quad (8)$$

where $\boldsymbol{\mu}_m$ lies in a $\hat{\boldsymbol{\mu}}_m$-centered ellipsoid shaped by $\boldsymbol{\Lambda}_m$ and bounded by $\gamma_m^\mu$.

iii) Covariance $\boldsymbol{\Sigma}_m$ for each component $m = 1, ..., M$:

$$\mathcal{C}_m^\Sigma = \left\{ \boldsymbol{\Sigma}_m \middle| \begin{array}{c} \|\boldsymbol{\Sigma}_m - \widehat{\boldsymbol{\Sigma}}_m\|_F \leq \gamma_m^\Sigma \\ \boldsymbol{\Sigma}_m \succ \boldsymbol{0} \end{array} \right\} \quad (9)$$

where $\|\cdot\|_F$ denotes the Frobenius norm, and $\boldsymbol{\Sigma}_m$ lies inside a $\widehat{\boldsymbol{\Sigma}}_m$-centered Frobenius-norm ball bounded by $\gamma_m^\Sigma$.

*C. Quantifying the Uncertainty of GMM Parameters*

Next, we provide two optional methods to quantify the uncertainty of $\pi_m$, $\boldsymbol{\mu}_m$ and $\boldsymbol{\Sigma}_m$ and determine the coefficients in the credible regions $\mathcal{C}^\pi$, $\mathcal{C}_m^\mu$ and $\mathcal{C}_m^\Sigma$.

*Method 1: A non-parametric bootstrap method* [26]

This method estimates the empirical distribution of a statistic by calculating the statistic on many bootstrap-resamples. This empirical distribution is used to construct the confidence region of the statistic. The bootstrap procedure is as follows [26]:

i) With all historical data of $\tilde{\boldsymbol{\xi}}$, use MLE to obtain the optimal parameters $M^*$ and $\boldsymbol{\Psi}^*$, where $M^*$ is chosen according to the Bayesian information criterion; then extract the maximum likelihood posterior group membership probability matrix $\mathbf{Z}^*$.

ii) Construct $N$ bootstrap resamples $\{Y_1, Y_2, ..., Y_N\}$ by drawing random observations with replacement from the original historical observations of $\tilde{\boldsymbol{\xi}}$.

iii) For each $Y_n$, initialize the MLE with $M^*$ and $\mathbf{Z}^*$; then conduct the MLE to obtain the corresponding optimal $\boldsymbol{\Psi}^{*n} = \{\pi_1^{*n}, \cdots, \pi_{M^*}^{*n}, \boldsymbol{\mu}_1^{*n}, \cdots, \boldsymbol{\mu}_{M^*}^{*n}, \boldsymbol{\Sigma}_1^{*n}, \cdots, \boldsymbol{\Sigma}_{M^*}^{*n}\}$.

With the obtained parameters $\boldsymbol{\Psi}^{*n}$, we can determine the coefficient of the credible regions in Section III-B.

For the credible region $\mathcal{C}^\pi$, we estimate

$$\bar{\pi}_m^\delta = \Gamma_{\frac{1+\delta}{2}}[\pi_m^{*1}, \cdots, \pi_m^{*N}] \quad (10a)$$

$$\underline{\pi}_m^\delta = \Gamma_{\frac{1-\delta}{2}}[\pi_m^{*1}, \cdots, \pi_m^{*N}] \tag{10b}$$

where $\Gamma_{\frac{1+\delta}{2}}[\cdots]$ and $\Gamma_{\frac{1-\delta}{2}}[\cdots]$ are the upper $\frac{1+\delta}{2}$ and $\frac{1-\delta}{2}$ quantiles of a data series, respectively.

For the credible region $\mathcal{C}_m^\mu$, we estimate

$$\hat{\boldsymbol{\mu}}_m = \frac{1}{N}\sum_{n=1}^N \boldsymbol{\mu}_m^{*n} \tag{11a}$$

$$\boldsymbol{\Lambda}_m = \frac{1}{N-1}\sum_{n=1}^N (\boldsymbol{\mu}_m^{*n} - \hat{\boldsymbol{\mu}}_m)(\boldsymbol{\mu}_m^{*n} - \hat{\boldsymbol{\mu}}_m)^T \tag{11b}$$

$$\gamma_m^\mu = \Gamma_\delta \left[\begin{array}{c}(\boldsymbol{\mu}_m^{*1} - \hat{\boldsymbol{\mu}}_m)^T \boldsymbol{\Lambda}_m^{-1}(\boldsymbol{\mu}_m^{*1} - \hat{\boldsymbol{\mu}}_m), \cdots, \\ (\boldsymbol{\mu}_m^{*N} - \hat{\boldsymbol{\mu}}_m)^T \boldsymbol{\Lambda}_m^{-1}(\boldsymbol{\mu}_m^{*N} - \hat{\boldsymbol{\mu}}_m)\end{array}\right] \tag{11c}$$

For the credible region $\mathcal{C}_m^\Sigma$, we estimate

$$\hat{\boldsymbol{\Sigma}}_m = \frac{1}{N}(\boldsymbol{\Sigma}_m^{*1} + \cdots + \boldsymbol{\Sigma}_m^{*N}) \tag{12a}$$

$$\gamma_m^\Sigma = \Gamma_\delta\left[\|\boldsymbol{\Sigma}_m^{*1} - \hat{\boldsymbol{\Sigma}}_m\|_F, \cdots, \|\boldsymbol{\Sigma}_m^{*N} - \hat{\boldsymbol{\Sigma}}_m\|_F\right] \tag{12b}$$

*Method 2: A method from a Bayesian viewpoint*

This method performs asymptotic inference on the estimator of $\boldsymbol{\Psi}$. More specific, under suitable regularity conditions and for large data sizes of $\tilde{\boldsymbol{\xi}}$, the posterior distribution of $\boldsymbol{\Psi}$ can be approximated by the following normal distribution and it does not depend on the prior distribution of $\boldsymbol{\Psi}$ (see Section 4 in [27]):

$$\phi\left(\boldsymbol{\Psi}^*, (\hat{\boldsymbol{I}}(\boldsymbol{\Psi}^*)J)^{-1}\right)$$

where $\boldsymbol{\Psi}^*$ is the MLE of $\boldsymbol{\Psi}$, $\hat{\boldsymbol{I}}(\boldsymbol{\Psi}^*)$ is the estimated Fisher information matrix evaluated at $\boldsymbol{\Psi}^*$ and it can be calculated based on the methods in [28], $J$ is the data size of $\tilde{\boldsymbol{\xi}}$. With this normal distribution, it is easy to determine the coefficients in the above credible regions and the details are omitted here.

The Bayesian method does not need the bootstrap resamples and is computationally very cheap. As the data size gets larger, the Bayesian method can be more accurate and is expected to offer a lower level of conservativeness associated with the credible regions. This is desirable since it helps make less-conservative decisions when more data information is at hand.

### D. Distributionally Robust CVaR-constrained OPF

After determining the credible regions of the GMM parameters, we can construct the following GMM-based ambiguity set to account for an ambiguous GMM distribution:

$$\mathcal{D} = \left\{\sum_{m=1}^{M^*} \pi_m \phi(\boldsymbol{\mu}_m, \boldsymbol{\Sigma}_m) \left| \begin{array}{l} \boldsymbol{\pi}:(\pi_1, \ldots, \pi_{M^*}) \in \mathcal{C}^\pi, \\ \boldsymbol{\mu}_m \in \mathcal{C}_m^\mu, \boldsymbol{\Sigma}_m \in \mathcal{C}_m^\Sigma, \\ \forall m = 1, \ldots, M^* \end{array}\right.\right\} \tag{13}$$

With the ambiguity set in (13), we can consider the following distributionally robust CVaR (DR-CVaR) constraints:

$$\max_{\mathbb{P}\in\mathcal{D}} \mathrm{CVaR}_{1-\beta}\left(\boldsymbol{y}_k^T \tilde{\boldsymbol{\xi}}\right) \leq h_k(\boldsymbol{x}), \forall k=1,\cdots,K \tag{14}$$

where the worst-case CVaR of $\boldsymbol{y}_k^T \tilde{\boldsymbol{\xi}}$ is no higher than $h_k(\boldsymbol{x})$.

Now, we formulate our DR-CVaR-constrained OPF as (P2):

$$\textbf{(P2)}: \quad \min_{\boldsymbol{x}}(2a), \text{ s.t.: } (2b)\text{-}(2f), (2i), (14). \tag{15}$$

## IV. SOLUTION METHODOLOGY

Next, we design an iterative cutting-plane algorithm to enforce the DR-CVaR constraints (14) and solve problem (P2).

### A. A Cutting-plane Method

To handle the DR-CVaR constraints (14), we will utilize two important and very desirable properties of CVaR:

i) since $\boldsymbol{y}_k^T \tilde{\boldsymbol{\xi}}$ is convex (affine) in $\boldsymbol{y}_k$, $\mathrm{CVaR}_{1-\beta}(\boldsymbol{y}_k^T \tilde{\boldsymbol{\xi}})$ is convex in $\boldsymbol{y}_k$ (see the Corollary 11 in [12]).

ii) it is easy to verify that $\boldsymbol{y}_k^T \tilde{\boldsymbol{\xi}}$ satisfies the Assumptions 1-3 in [29], so the gradients of $\mathrm{CVaR}_{1-\beta}(\boldsymbol{y}_k^T \tilde{\boldsymbol{\xi}})$ with respect to $\boldsymbol{y}_k$ can be calculated as the following conditional expectations (see Theorem 3.1 in [29]):

$$\nabla_{\boldsymbol{y}_{k,i}}\mathrm{CVaR}_{1-\beta}\left(\boldsymbol{y}_k^T \tilde{\boldsymbol{\xi}}\right) \\ = \mathrm{E}_{\mathbb{P}}\left[\tilde{\xi}_i \big| \boldsymbol{y}_k^T \tilde{\boldsymbol{\xi}} \geq \mathrm{VaR}_{1-\beta}\left(\boldsymbol{y}_k^T \tilde{\boldsymbol{\xi}}\right)\right], \forall i \tag{16}$$

where $\boldsymbol{y}_{k,i}$ denotes the $i_{th}$ element of $\boldsymbol{y}_k$, and $\mathrm{VaR}_{1-\beta}(\boldsymbol{y}_k^T \tilde{\boldsymbol{\xi}})$ is the value-at-risk of $\boldsymbol{y}_k^T \tilde{\boldsymbol{\xi}}$ at the probability level $1-\beta$.

In [29], a Monte-Carlo sampling method is used to estimate the VaR, CVaR and its gradients in (16). However, this method is computationally inefficient, and it is difficult to account for the distributional ambiguity of $\tilde{\boldsymbol{\xi}}$ within such method.

Next, we will show that, thanks to the use of GMM, a precise and scalable method can be applied to calculate the VaR, CVaR and gradients in (16) while considering the distributional ambiguity of $\tilde{\boldsymbol{\xi}}$ as modelled by (13).

*Step 1*: identify the worst-case $\boldsymbol{\mu}_m$ and $\boldsymbol{\Sigma}_m$.

For a given $\boldsymbol{y}_k^*$, the corresponding worst-case CVaR of $\boldsymbol{y}_k^{*T}\tilde{\boldsymbol{\xi}}$ can be reformulated as (17) by combining (4), (13) and (14):

$$\max_{\mathbb{P}\in\mathcal{D}} \mathrm{CVaR}_{1-\beta}\left(\boldsymbol{y}_k^{*T}\tilde{\boldsymbol{\xi}}\right)$$
$$= \max_{\mathbb{P}\in\mathcal{D}} \min_{t\in\mathbb{R}}\left\{t + \frac{1}{\beta}\mathrm{E}_{\mathbb{P}}\left[\boldsymbol{y}_k^{*T}\tilde{\boldsymbol{\xi}} - t\right]^+\right\}$$
$$= \max_{\substack{\boldsymbol{\pi}\in\mathcal{C}^\pi, \boldsymbol{\mu}_m\in\mathcal{C}_m^\mu,\\ \boldsymbol{\Sigma}_m\in\mathcal{C}_m^\Sigma}}\left\{\min_{t\in\mathbb{R}}\left[t + \frac{1}{\beta}\sum_{m=1}^{M^*}\pi_m Q_m\right]\right\} \tag{17}$$

In (17), the auxiliary variable $Q_m$ is an integral under a multivariate normal distribution

$$Q_m = \int_{\tilde{\boldsymbol{\xi}}\in\mathbb{R}^W}\left[\boldsymbol{y}_k^{*T}\tilde{\boldsymbol{\xi}} - t\right]^+ \phi(\tilde{\boldsymbol{\xi}}|\boldsymbol{\mu}_m, \boldsymbol{\Sigma}_m)d\tilde{\boldsymbol{\xi}}$$

and it has an analytical form

$$Q_m = \bar{\sigma}_m^2 \phi_m(t|\bar{\mu}_m, \bar{\sigma}_m) + (\bar{\mu}_m - t)[1 - \Phi_m(t|\bar{\mu}_m, \bar{\sigma}_m)] \tag{18}$$

where $\phi_m$ and $\Phi_m$ are respectively the PDF and cumulative distribution function (CDF) of the normal distribution with mean $\bar{\mu}_m = \boldsymbol{y}_k^{*T}\boldsymbol{\mu}_m$ and standard deviation $\bar{\sigma}_m = \sqrt{\boldsymbol{y}_k^{*T}\boldsymbol{\Sigma}_m\boldsymbol{y}_k^*}$.

It can be verified that $Q_m$ is monotonically increasing with $\bar{\mu}_m$ and $\bar{\sigma}_m$ for any $t$. Thus, the optimal solution of (17) corresponds to the maximum values of $\bar{\mu}_m$ and $\bar{\sigma}_m$.

The maximum $\bar{\mu}_m$ and $\bar{\sigma}_m$ can be obtained by identifying the worst-case $\boldsymbol{\mu}_m \in \mathcal{C}_m^\mu$ and worst-case $\boldsymbol{\Sigma}_m \in \mathcal{C}_m^\Sigma$. Due to the special structures of the $\mathcal{C}_m^\mu$ and $\mathcal{C}_m^\Sigma$ in (8)-(9), the worst-case $\boldsymbol{\mu}_m$ and $\boldsymbol{\Sigma}_m$ have closed-form solutions:

$$\bar{\mu}_m^{max} = \boldsymbol{y}_k^{*T} \boldsymbol{\mu}_m^{wc} \tag{19a}$$

$$\boldsymbol{\mu}_m^{wc} = \hat{\boldsymbol{\mu}}_m + \sqrt{\frac{\gamma_m^\mu}{\boldsymbol{y}_k^{*T} \boldsymbol{\Lambda}_m \boldsymbol{y}_k^*}} \boldsymbol{\Lambda}_m \boldsymbol{y}_k^* \tag{19b}$$

$$\bar{\sigma}_m^{max} = \sqrt{\boldsymbol{y}_k^{*T} \boldsymbol{\Sigma}_m^{wc} \boldsymbol{y}_k^*} \tag{19c}$$

$$\boldsymbol{\Sigma}_m^{wc} = \widehat{\boldsymbol{\Sigma}}_m + \gamma_m^\Sigma \frac{\boldsymbol{y}_k^* \boldsymbol{y}_k^{*T}}{\|\boldsymbol{y}_k^*\|_2^2} \tag{19d}$$

where the superscript $wc$ means 'the worst-case'.

*Step 2*: identify the worst-case $\boldsymbol{\pi}$ and the corresponding $\text{VaR}_{1-\beta}^{wc}(\boldsymbol{y}_k^{*T}\tilde{\boldsymbol{\xi}})$.

It is easy to verify that $Q_m$ is convex in $t$, so (17) can be rewritten as (20) according to the max-min theorem [30]:

$$\min_{t \in \mathbb{R}} \left[ t + \frac{1}{\beta} \max_{\boldsymbol{\pi} \in \mathcal{C}^\pi} \sum_{m=1}^{M^*} \pi_m Q_m|_{\bar{\mu}_m^{max}, \bar{\sigma}_m^{max}} \right] \tag{20}$$

To simplify expressions, let $T(t)$ be the function inside the square bracket in (20), and so (20) is equal to

$$\min_{t \in \mathbb{R}} T(t) \tag{21}$$

Due to the convexity of $Q_m$ in $t$, it is easy to proof that $T(t)$ is also convex in $t$. Also, problem (21) involves only one decision variable $t$, and $T(t)$ could be non-smooth. These observations motivate us to design a bisection search method to solve (21):

---

**Algorithm 1**: A bisection search method to solve (21)

1. Receive $\bar{\mu}_m^{max}$ and $\bar{\sigma}_m^{max}, \forall m$.
2. Let $\check{t}_m = \bar{\mu}_m^{max} + \Phi^{-1}(1-\beta)\bar{\sigma}_m^{max}, \forall m$, where $\Phi^{-1}(1-\beta)$ is the inverse of the CDF of standard normal distribution evaluated at $1-\beta$.
3. Let $\underline{t} = \min_m \check{t}_m$, $\bar{t} = \max_m \check{t}_m$.
4. **while** $\bar{t} - \underline{t} > 10^{-5}$
       $t^o = (\underline{t} + \bar{t})/2$;
       Calculate $Q_m(t^o)|_{\bar{\mu}_m^{max}, \bar{\sigma}_m^{max}}, \forall m$, according to (18);
       $\boldsymbol{\pi}^o = \text{SortM}(\mathcal{C}^\pi, Q_m(t^o)|_{\bar{\mu}_m^{max}, \bar{\sigma}_m^{max}}, \forall m)$;
       Fix the $\boldsymbol{\pi}$ in $T(t)$ as $\boldsymbol{\pi}^o$, calculate the gradient of $T(t)$ at $t^o$:
       $\nabla_t T(t)|_{\boldsymbol{\pi}^o, \bar{\mu}_m^{max}, \bar{\sigma}_m^{max}} = 1 - \frac{1}{\beta} + \frac{1}{\beta}\sum_{m=1}^{M^*}[\pi_m^o \Phi_m(t^o|\bar{\mu}_m^{max}, \bar{\sigma}_m^{max})]$.
       **if** $\nabla_t T(t)|_{\boldsymbol{\pi}^o, \bar{\mu}_m^{max}, \bar{\sigma}_m^{max}} < 0$, **then**
           $\underline{t} = t^o$;
       **else**
           $\bar{t} = t^o$;
       **end if**
   **end while**
5. Output $\boldsymbol{\pi}^{wc} = \boldsymbol{\pi}^o$ and the corresponding $\text{VaR}_{1-\beta}^{wc}(\boldsymbol{y}_k^T\tilde{\boldsymbol{\xi}}) = t^o$.

---

where the structure of $\mathcal{C}^\pi$ allows us to use the sorting method in [31] (i.e., SortM) to identify the $\boldsymbol{\pi} \in \mathcal{C}^\pi$ that corresponds to the maximum $\sum_{m=1}^{M^*}\pi_m Q_m(t^o)|_{\bar{\mu}_m^{max}, \bar{\sigma}_m^{max}}$ (i.e., $\boldsymbol{\pi}^o$).

*Step 3*: develop the cutting-plane of $\max_{\mathbb{P}\in\mathcal{D}}\text{CVaR}_{1-\beta}(\boldsymbol{y}_k^T\tilde{\boldsymbol{\xi}})$

With the identified $\boldsymbol{\mu}_m^{wc}$, $\boldsymbol{\Sigma}_m^{wc}$ and $\boldsymbol{\pi}^{wc}$, the worst-case $\mathbb{P} \in \mathcal{D}$ is determined and let $\text{CVaR}_{1-\beta}^{wc}(\boldsymbol{y}_k^T\tilde{\boldsymbol{\xi}})$ be the CVaR corresponding to such $\mathbb{P}$. For the given $\boldsymbol{y}_k^*$, $\text{CVaR}_{1-\beta}^{wc}(\boldsymbol{y}_k^{*T}\tilde{\boldsymbol{\xi}})$ can be calculated by substituting $\bar{\mu}_m^{max}$, $\bar{\sigma}_m^{max}$, $\boldsymbol{\pi}^{wc}$ and $t = \text{VaR}_{1-\beta}^{wc}(\boldsymbol{y}_k^{*T}\tilde{\boldsymbol{\xi}})$ into (17)-(18). The gradients of $\text{CVaR}_{1-\beta}^{wc}(\boldsymbol{y}_k^T\tilde{\boldsymbol{\xi}})$ at $\boldsymbol{y}_k^*$ can be obtained according to (16) with the analytical forms:

$$\begin{aligned}
&\nabla_{\boldsymbol{y}_{k,i}}\text{CVaR}_{1-\beta}^{wc}(\boldsymbol{y}_k^{*T}\tilde{\boldsymbol{\xi}}) \\
&= \frac{1}{\beta}\sum_{m=1}^{M^*}\pi_m^{wc}\left\{\boldsymbol{\mu}_{m,i}^{wc}\left[1 - \Phi_m\left(\text{VaR}_{1-\beta}^{wc}(\boldsymbol{y}_k^{*T}\tilde{\boldsymbol{\xi}})|\bar{\mu}_m^{max},\bar{\sigma}_m^{max}\right)\right]\right. \\
&\left. + \boldsymbol{\Sigma}_{m,(i,\cdots)}^{wc}\boldsymbol{y}^{*T}\phi_m\left(\text{VaR}_{1-\beta}^{wc}(\boldsymbol{y}_k^{*T}\tilde{\boldsymbol{\xi}})|\bar{\mu}_m^{max},\bar{\sigma}_m^{max}\right)\right\}, \forall i
\end{aligned} \tag{22}$$

It is easy to verify that $\max_{\mathbb{P}\in\mathcal{D}}\text{CVaR}_{1-\beta}(\boldsymbol{y}_k^T\tilde{\boldsymbol{\xi}})$ is also convex in $\boldsymbol{y}_k$, and the value and gradients of $\max_{\mathbb{P}\in\mathcal{D}}\text{CVaR}_{1-\beta}(\boldsymbol{y}_k^T\tilde{\boldsymbol{\xi}})$ at $\boldsymbol{y}_k^*$ are the same as those of $\text{CVaR}_{1-\beta}^{wc}(\boldsymbol{y}_k^T\tilde{\boldsymbol{\xi}})$. Thus, we can use the cutting-plane supporting $\text{CVaR}_{1-\beta}^{wc}(\boldsymbol{y}_k^T\tilde{\boldsymbol{\xi}})$ at $\boldsymbol{y}_k^*$ to linearize the $k_{th}$ DR-CVaR constraint in (14). That is

$$\text{CVaR}_{1-\beta}^{wc}(\boldsymbol{y}_k^{*T}\tilde{\boldsymbol{\xi}}) + \sum_i\left[\nabla_{\boldsymbol{y}_{k,i}}\text{CVaR}_{1-\beta}^{wc}(\boldsymbol{y}_k^{*T}\tilde{\boldsymbol{\xi}})(\boldsymbol{y}_{k,i} - \boldsymbol{y}_{k,i}^*)\right] \leq h_k(\boldsymbol{x}) \tag{23}$$

### B. Iterative Solution Algorithm for Problem (P2)

With the cutting-plane-based constraint (23), we can develop an iterative solution algorithm to gradually enforce the DR-CVaR constraints in (14) and solve the problem (P2) in (15).

---

**Algorithm 2**: Iterative solution algorithm for problem (P2)

1. Solve the following problem (**P3**), which is the relaxation of (P2):
(**P3**): $\{\min_{\boldsymbol{x}}(2a)$, s.t.: (2b)-(2f), (2i) and the cutting-plane-based constraints which have been generated $\}$
2. Obtain the optimal solution $\boldsymbol{x}^*$, and form $\boldsymbol{y}_k^*, \forall k = 1, \cdots, K$.
3. **for** $k = 1, \cdots, K$, **do**
       Use the Steps 1-3 in Section IV-A to calculate $\text{CVaR}_{1-\beta}^{wc}(\boldsymbol{y}_k^{*T}\tilde{\boldsymbol{\xi}})$.
       **if** $\text{CVaR}_{1-\beta}^{wc}(\boldsymbol{y}_k^{*T}\tilde{\boldsymbol{\xi}}) > h_k(\boldsymbol{x}^*)$ **then**
           Generate the cutting-plane-based constraint (23).
       **end if**
**end for**
4. **if** $\text{CVaR}_{1-\beta}^{wc}(\boldsymbol{y}_k^{*T}\tilde{\boldsymbol{\xi}}) \leq h_k(\boldsymbol{x}^*), \forall k$: terminate; **Otherwise**, go to Step 1.

---

We also have one enhancement to accelerate Algorithm 2. The upward-reserve DR-CVaR constraints are rewritten as:

$$\alpha_g \max_{\mathbb{P}\in\mathcal{D}}\text{CVaR}_{1-\beta}(-\boldsymbol{1}^T\tilde{\boldsymbol{\xi}}) \leq R_g^{up}, \forall g \tag{24}$$

We can first calculate the worst-case CVaR of $-\boldsymbol{1}^T\tilde{\boldsymbol{\xi}}$ through the Steps 1-3 in Section IV-A (the result is denoted as $\Theta^{UP}$) and use (25) to replace the original constraints in (24).

$$\alpha_g \Theta^{UP} \leq R_g^{up}, \forall g \tag{25}$$

Similarly, after calculating the worst-case CVaR of $\boldsymbol{1}^T\tilde{\boldsymbol{\xi}}$ evaluated at $1-\beta$ (the result is $\Theta^{DN}$), we can use (26) to replace the downward-reserve DR-CVaR constraints in (P2).

$$\alpha_g \Theta^{DN} \leq R_g^{dn}, \forall g \tag{26}$$

Now, the reserve-related DR-CVaR constraints in (P2) are replaced by linear constraints (25)-(26), and we only need to apply the cutting-planes to branch-flow DR-CVaR constraints.

## C. Remarks on the Solution Methodology

1) The whole methodology does not rely on any sampling procedure. First, for the evaluation of the worst-case CVaR, the Steps 1-3 in Section IV-A can be quickly implemented since they do not require any sample-based simulations but only involve the bisection search and some analytical solutions. Second, for the solving of the proposed OPF model, cutting-planes are used to enforce the DR-CVaR constraints. Compared to the traditional methods with sample-based approximation of CVaR (e.g., [24]), the cutting-plane method is computationally more desirable since it makes the scale of the proposed OPF model irrelevant to the size of historical samples.

2) The structures of $\mathcal{C}^\pi$, $\mathcal{C}_m^\mu$, $\mathcal{C}_m^\Sigma$ in (7)-(9) are crucial for the scalability of the Steps 1-3 in Section IV-A: a) the structures of $\mathcal{C}_m^\mu$ and $\mathcal{C}_m^\Sigma$ for the analytical solutions of $\overline{\mu}_m^{max}$ and $\overline{\sigma}_m^{max}$; b) the structure of $\mathcal{C}^\pi$ for the fast implementation of Algorithm 1.

3) Our experiments show that usually a very limited number of cutting-planes are actually active in each iteration, so the size of problem (P3) increases slowly with the iteration.

## V. CASE STUDIES

The proposed OPF model is tested on a modified IEEE 2736-bus system, with the original data from [32]. The system has 2736 buses, 289 controllable units, 3504 transmission lines and 10 wind farms. The total capacities of the controllable units and wind farms are 28.88 GW and 3.08 GW, respectively. The flow limits of all transmission lines are reduced by 20% to create transfer congestions. All numerical tests are carried out on a desktop with Intel Core i7-8700 CPU and 16GB RAM.

### A. Construction of the Proposed Ambiguity set

In this paper, the underlying marginal distributions of the forecast errors at all wind farms and the associated correlations are extracted from Wind Integration National Dataset Toolkit [33]. Based on the marginal distributions and correlations, the Nataf method in [34] is used to generate some forecast error samples, which forms the historical dataset available to system operators. With the samples as the input, the bootstrap procedure in Section III-C is used to construct the proposed ambiguity set. The number of the bootstrap resamples is 2000. The confidence level $\delta$ in (10)-(12) is 95%.

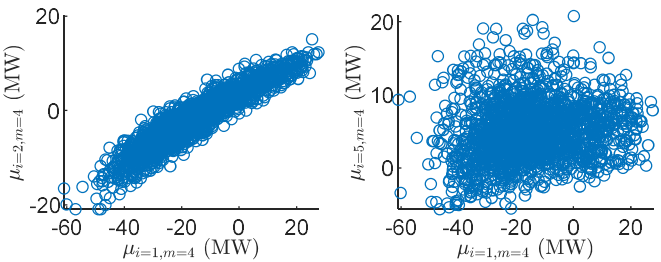

Fig. 1. (left) Bootstrap mean values of the forecast errors at wind farms 1 and 2; (right) Bootstrap mean values of the forecast errors at wind farms 1 and 5.

As an example, 4000 forecast error samples are generated as the input. The number of the Gaussian components given by the bootstrap method is six. For the 4th Gaussian component, the associated bootstrap means of the forecast errors at wind farms 1, 2 and 5 are plotted in Fig. 1. It is seen that there is a correlation between the means for wind farm 1 and 2. This is speculated to be related to the high dependency between the forecast errors themselves (the correlation coefficient of the two forecast errors is around 0.75). As a comparison, the means for wind farms 1 and 5 have no obvious dependency since the corresponding forecast errors are weakly correlated. The potential correlation between the means is another reason why we use ellipsoids as the credible regions of the means (the first reason can be seen in Section IV-C), considering ellipsoids have a good capability of modelling correlated factors [35].

### B. Comparison of Different CVaR-constrained OPF Models

The proposed OPF model, denoted by **DG-OPF**, is compared with three other OPF models with CVaR constraints:

i) **NA-OPF**: the GMM with fixed parameters is used and no distributional ambiguity is considered. Here, the GMM fitted at the first step of the bootstrap procedure is used.

ii) **M-OPF**: The traditional moment-based ambiguity set is used, which utilizes only the estimated mean $\hat{\boldsymbol{\mu}}$ and covariance $\widehat{\Sigma}$ of the forecast error (see its structure in [3], [4]). In this case, the worst-case CVaR of $\boldsymbol{y}_k^T \tilde{\boldsymbol{\xi}}$ has a simple form [21]:

$$\boldsymbol{y}_k^T \hat{\boldsymbol{\mu}} + \sqrt{\frac{1-\beta}{\beta}} \sqrt{\boldsymbol{y}_k^T \widehat{\Sigma} \boldsymbol{y}_k} \qquad (27)$$

iii) **U-OPF**: This model applies not only the mean and covariance information but also the unimodality knowledge to construct its ambiguity set. The model can be formulated as a semi-definite programming problem (see more details in [22]).

DG-OPF, NA-OPF and M-OPF are solved by GUROBI while U-OPF is solved by MOSEK.

We consider two sample sizes: 200 and 4000. The risk levels $\beta$ for reserve and branch flow-CVaR constraints are set to 2% and 4%, respectively. For each sample size, we do the following procedure ten times: a) generate samples; b) solve all the four models; c) re-generate $10^6$ forecast error scenarios using the Nataf method; d) for each model, keep the optimal operating strategy fixed, and do the out-of-sample test (to check if all CVaR constraints are satisfied under the $10^6$ scenarios).

For each OPF model, the average, maximum and minimum cost results are listed in Table I. Meanwhile, for each model, the worst-case violations of various CVaR constraints observed in the out-of-sample tests are reported in Table II. It is seen from Table I and II that NA-OPF has the lowest cost but fails to meet the reliability requirements (the violated CVaR constraints are marked by boldface type). The reason is that NA-OPF makes decisions based on a deterministic GMM distribution, which is estimated from a small dataset with limited distributional information. This GMM could deviate from the underlying true distribution, and the NA-OPF dispatch may become unsafe. In contrast to NA-OPF, each of the other three models accounts for not a single but a family of possible true distributions. This needs additional cost but can hedge against the ambiguity in the true distribution and yield more robust operational strategies.

For DG, M and U-OPF models, they all satisfy the reliability requirements, but this is at the expense of different costs. As shown in Table I, M-OPF needs the highest total cost and the inclusion of the unimodality information in U-OPF decreases the required cost. Compared to M- and U-OPF, DG-OPF needs much lower cost. To better present this point, the additional costs of the three models used to hedge against the distributional ambiguity (i.e., the additional costs in relative to the total cost of NA-OPF) are listed in Table III. Under the data size 200, M and U-OPF need additional 7.97% and 6.74% costs in average, respectively, while such percentage is reduced to 2.21% in DG-OPF. Similar results are observed for the data size 4000.

Table I. Total Costs for Different OPF Models

| Data size | Total cost ($10^6$\$) | NA-OPF | DG-OPF | M-OPF | U-OPF |
|---|---|---|---|---|---|
| 200 | avg | 1.0792 | 1.1031 | 1.1653 | 1.1520 |
|  | max | 1.0831 | 1.1189 | 1.1727 | 1.1574 |
|  | min | 1.0747 | 1.0932 | 1.1511 | 1.1399 |
| 4000 | avg | 1.0790 | 1.1021 | 1.1627 | 1.1495 |
|  | max | 1.0799 | 1.1162 | 1.1645 | 1.1511 |
|  | min | 1.0783 | 1.0941 | 1.1597 | 1.1469 |

Table II. Worst-case Violations of Various CVaR-constraints in the Out-of-sample Tests (MW)

| Data size | Type of CVaR-constraint | NA-OPF | DG-OPF | M-OPF | U-OPF |
|---|---|---|---|---|---|
| 200 | Upward reserve | **115.20*** | -74.64 | -708.72 | -617.45 |
|  | Downward reserve | -4.45 | -217.77 | -852.21 | -696.00 |
|  | Branch Flow | **3.66** | -0.21 | -1.12 | -0.94 |
| 4000 | Upward reserve | **35.26** | -133.12 | -796.42 | -698.88 |
|  | Downward reserve | -9.53 | -158.47 | -927.03 | -759.98 |
|  | Branch Flow | **0.47** | -0.25 | -1.22 | -1.04 |

* a positive value indicates the associated CVaR constraint is violated and it is the magnitude of the violation; a negative value indicates the CVaR constraint is not violated and it is the safety gap between the CVaR and operational limit.

Table III. Comparison of DG, M and U-OPF Models in Terms of the Additional Cost to Hedge Against the Distributional Ambiguity

| Data size | Additional cost (%) | DG-OPF | M-OPF | U-OPF |
|---|---|---|---|---|
| 200 | avg | 2.2080 | 7.9743 | 6.7385 |
|  | max | 3.3471 | 8.3242 | 7.1282 |
|  | min | 1.7224 | 7.1147 | 6.0642 |
| 4000 | avg | 2.1374 | 7.7492 | 6.5295 |
|  | max | 3.3923 | 7.9113 | 6.6655 |
|  | min | 1.4215 | 7.5450 | 6.3626 |

The difference in the conservativeness of DG, M and U-OPF can be also observed in Table IV, which lists the amounts of the upward reserves procured by these models. In average, the amount for M-OPF is more than 70% higher than that for DG-OPF. Though the amount for M-OPF guarantees no violations of the upward reserve-related CVaR constraints, it could be unnecessarily high.

Table IV. Upward Reserves Procured by Different OPF Models

| Data size | Upward Reserve (MW) | NA-OPF | DG-OPF | M-OPF | U-OPF |
|---|---|---|---|---|---|
| 200 | avg | 592.83 | 834.98 | 1496.72 | 1393.93 |
|  | max | 688.28 | 1072.74 | 1602.50 | 1481.33 |
|  | min | 525.92 | 715.76 | 1349.84 | 1258.56 |
| 4000 | avg | 625.30 | 848.82 | 1463.64 | 1365.99 |
|  | max | 645.28 | 941.57 | 1482.60 | 1382.59 |
|  | min | 605.85 | 774.24 | 1437.54 | 1339.99 |

We also compare the worst-case distributions in M, U and DG-OPF models. For the upward reserve CVaR constraints, the associated worst-case distributions of the total wind power forecast error are plotted in Fig. 2. The empirical histogram of the total forecast error formed by the $10^6$ scenarios for out-of-sample-tests and the upward reserves procured by the three models are also plotted. The worst-case distribution of M-OPF is a two-supported discrete distribution, and the left support has a total probability mass of 2% and is far from the zero point, which causes the procurement of the large amount of upward reserve; In the U-OPF with the unimodality information, the worst-case distribution consists of two uniform distributions, and the left one has a total probability mass of 2.89%, and the occurrence probability of very large downward forecast errors is reduced; For DG-OPF, its worst-case distribution is still a continuous GMM and it has a shorter stretch at the left-hand tail compared to the cases in M and U-OPF, and thus the occurrence probability of very large downward forecast errors is the lowest.

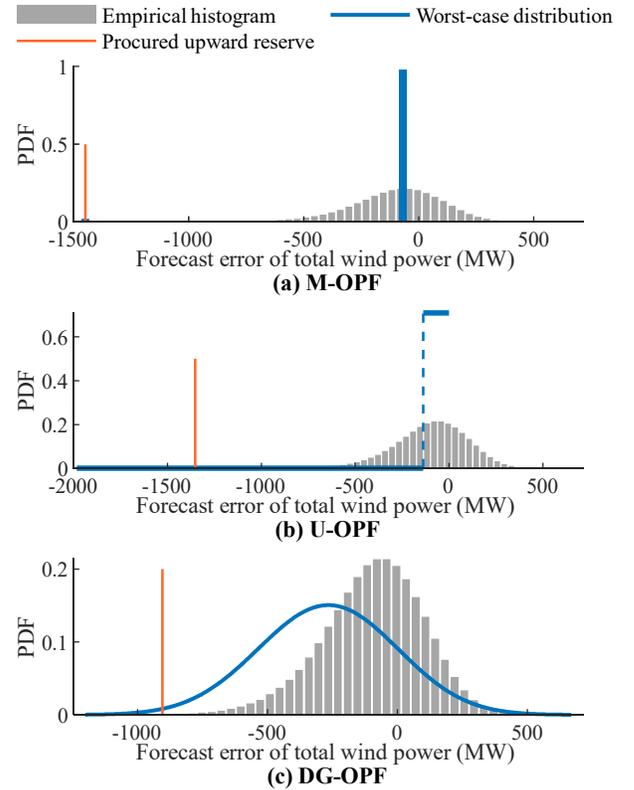

Fig. 2. Worst-case distributions of the total wind power forecast error associated with M, U and DG-OPF (the results correspond to the data size 4000).

Table V. Computational Performances of DG, M and U-OPF Models

| Data size | Total CPU time (s) | DG-OPF | M-OPF | U-OPF |
|---|---|---|---|---|
| 200 | avg | 3.26 | 1.95 | 8.75 |
|  | max | 3.56 | 2.74 | 12.10 |
|  | min | 3.05 | 1.56 | 5.14 |
| 4000 | avg | 3.52 | 2.20 | 8.87 |
|  | max | 3.87 | 2.64 | 9.04 |
|  | min | 3.08 | 1.54 | 8.78 |

Table V lists the solving time of DG, M, and U-OPF. As can be seen, the solving time of DG-OPF is longer than that of M-OPF but shorter than that of U-OPF. The reasons include that DG-OPF has more complex reformulations of DR-CVaR constraints compared to M-OPF, and DG-OPF involves only linear programs while U-OPF is a semi-definite model. Still, DG-OPF can be solved within seconds on this 2736-bus system. This proves the usefulness of DG-OPF in short-term dispatch (e.g., 5 minutes ahead) on large networks. The solving time of DG-OPF also changes slightly as the data size increases. This is expected since the scale of DG-OPF is irrelated to the data size. Moreover, DG-OPF uses averagely only 37 ms to evaluate the worst-case CVaRs for totally 7008 branch flow constraints. This shows the scalability of the Steps 1-3 in Section IV-A.

In summary, compared to M and U-OPF, DG-OPF uses more economical dispatch to satisfy the same reliability requirements while still maintaining a high computational efficiency.

## VI. Conclusion

This paper proposes a GMM-based distributionally robust OPF model with CVaR constraints. A new ambiguity set based on uncertain GMM parameters is constructed, and a scalable cutting-plane-based solution methodology is designed. Case studies on a modified 2736-bus system show that the proposed GMM-based ambiguity set is capable of effectively hedging against the distributional ambiguity of wind power forecast error while offering less conservative dispatch than the moment and unimodality-based ambiguity sets. The case studies also reveal the scalability of the proposed OPF model, validating its potential application for short-term dispatch on large networks.